%% file: main.tex
\documentclass[sigconf, nonacm, pdfa]{acmart}

\usepackage{listings}
\usepackage{semantic}
\usepackage[capitalise, noabbrev]{cleveref}
\usepackage{multirow}
\usepackage{enumitem}
\usepackage{booktabs}
\usepackage{pgfplots}
\usepackage{framed}

\pgfplotsset{
  width=8cm,
  compat=1.9,
  /pgfplots/ybar legend/.style={
  /pgfplots/legend image code/.code={%
      \draw[##1,/tikz/.cd,yshift=-0.25em]
      (0cm,0cm) rectangle (3pt,0.8em);},
   },
}

\newcommand\graphalg{\textsc{GraphAlg}}
\newcommand\matlang{MATLANG}

\definecolor{pred}{rgb}{0.5,0,0.5}
\definecolor{pgreen}{rgb}{0,0.5,0}

\lstdefinelanguage{GraphAlg}{
  morekeywords={func, for, in, until, return}, % Control Flow
  morekeywords={Matrix, Vector, bool, int, real, trop\_int, trop\_real, trop\_max\_int}, % Types
  morekeywords={T, diag, apply, select, tril, triu, reduceRows, reduceCols, reduce, pickAny, zero, one, cast}, % Built-in functions.
  morekeywords={nrows, cols, nvals}, % Properties.
  keywordstyle=\color{blue}\bfseries,
  identifierstyle=\color{black},
  sensitive=false,
  comment=[l]{//},
  commentstyle=\color{purple}\ttfamily,
}

\lstdefinelanguage{GraphAlgInCypher}{
  morekeywords={func, for, in, until, return, WITH, ALGORITHM, CALL, AS}, % Control Flow
  morekeywords={Matrix, Vector, bool, int, real, trop\_int, trop\_real, trop\_max\_int}, % Types
  morekeywords={T, diag, apply, select, tril, triu, reduceRows, reduceCols, reduce, pickAny, zero, one}, % Built-in functions.
  morekeywords={nrows, cols, nvals}, % Properties.
  keywordstyle=\color{blue}\bfseries,
  identifierstyle=\color{black},
  sensitive=false,
  comment=[l]{//},
  commentstyle=\color{purple}\ttfamily,
}

\lstdefinelanguage{SQLInPython}{
  morekeywords={SELECT, SUM, FROM, JOIN, USING, CREATE, TABLE, AS, SELECT, LEFT, ON, GROUP, BY, DROP}, % SQL keywords 
  keywordstyle=\color{blue}\bfseries,
  identifierstyle=\color{black},
  sensitive=false,
  comment=[l]{//},
  commentstyle=\color{purple}\ttfamily,
}

\begin{document}
\title{Algorithm Support for Graph Databases, Done Right}

\author{Daan de Graaf}
\affiliation{%
  \institution{Eindhoven University of Technology}
  \city{Eindhoven}
  \country{Netherlands}
}
\email{d.j.a.d.graaf@tue.nl}

\author{Robert Brijder}
\affiliation{%
  \institution{Eindhoven University of Technology}
  \city{Eindhoven}
  \country{Netherlands}
}
\email{r.brijder@tue.nl}

\author{Soham Chakraborty}
\affiliation{%
  \institution{Delft University of Technology}
  \city{Delft}
  \country{Netherlands}
}
\email{s.s.chakraborty@tudelft.nl}

\author{George Fletcher}
\affiliation{%
  \institution{Eindhoven University of Technology}
  \city{Eindhoven}
  \country{Netherlands}
}
\email{g.h.l.fletcher@tue.nl}

\author{Bram van de Wall}
\affiliation{%
  \institution{Eindhoven University of Technology}
  \city{Eindhoven}
  \country{Netherlands}
}
\email{a.a.g.v.d.wall@tue.nl}

\author{Nikolay Yakovets}
\affiliation{%
  \institution{Eindhoven University of Technology}
  \city{Eindhoven}
  \country{Netherlands}
}
\email{hush@tue.nl}

\begin{abstract}
  Graph database query languages cannot express algorithms like PageRank, forcing costly data wrangling, while existing solutions such as algorithm libraries, vertex-centric APIs, and recursive CTEs lack the necessary combination of expressiveness, performance, and usability.
  We present \graphalg{}: a domain-specific language for graph algorithms that compiles to relational algebra, enabling seamless integration with query processing pipelines.
  Built on linear algebra foundations, \graphalg{} provides intuitive matrix operations that are amenable to aggressive optimization including sparsity analysis, loop-invariant code motion, and in-place aggregation.
  Our implementation in AvantGraph demonstrates significant code complexity reduction compared to SQL/Python and Pregel while achieving excellent performance on LDBC Graphalytics benchmarks.
  \graphalg{} establishes that graph databases can serve as unified platforms for both queries and analytics.
\end{abstract}

\maketitle

  \begin{framed}
  \section*{Artifacts}
  Source code for the \graphalg{} compiler is freely available at \href{https://github.com/wildarch/graphalg}{github.com/wildarch/graphalg}.

  \vspace{18pt}

  \noindent Further documentation can be found at \href{https://wildarch.dev/graphalg/}{wildarch.dev/graphalg}.
  This includes an interactive tutorial on writing programs in GraphAlg, as well as documentation on the internals of the compiler.
  \end{framed}

\begin{table*}[t]
  \caption{
    An overview of existing approaches to graph analytics support in database systems.
    Algorithms library, Pregel API and algorithm DSL are solutions tailored to the problem of graph analytics.
    The other options are more generic tools available in relational database systems.
    The listed systems are a representative subset of popular database systems rather than an exhaustive list.
  }
  \label{tab:existing-approaches}
  \begin{tabular}{l|l|l}
    \toprule
    \textbf{Approach}      & \textbf{Key Problems}                  & \textbf{Available in}                         \\
    \midrule
    Algorithms Library     & Fixed set of Algorithms                & Neo4j, ArangoDB                               \\
    Pregel API             & Performance issues, No optimization    & Neo4j                                         \\
    User-defined Operators & Unsafe, No optimization                & Umbra                                         \\
    Recursive CTEs         & Difficult to write, Performance issues & DuckDB, Umbra, PostgreSQL, Oracle, SQL Server \\
    Procedural SQL         & Overhead, Limited optimization         & PostgreSQL, Oracle, SQL Server                \\
    Algorithm DSL          & Only proprietary implementations       & TigerGraph, Oracle PGX                        \\
    \bottomrule
  \end{tabular}
\end{table*}

\section{Introduction}
Consider a data scientist analyzing citation networks to identify influential papers using PageRank, or a fraud analyst running community detection algorithms on transaction graphs.
Despite graph databases being the natural home for such data, these users cannot express these fundamental algorithms in Cypher~\cite{francis_cypher_2018} or GQL~\cite{iso_information_2024}, the very languages designed for graph data!
Instead, they must export gigabytes of data, wrestle with format conversions, and maintain duplicate copies that quickly become stale.
This disconnect between graph storage and graph analytics represents a significant gap in current database systems.

While some databases have attempted solutions, e.g., algorithm libraries (ArangoDB), a Pregel API (Neo4J), or recursive CTEs (PostgreSQL), each approach has significant limitations (see \cref{tab:existing-approaches}).
Libraries offer a fixed set of algorithms that rarely match exact requirements.
Vertex-centric APIs suffer from poor performance due to excessive message passing.
Recursive CTEs are notoriously difficult to write and optimize.
The result is that users continue to rely on external tools, negating the very benefits of integrated data management.

This paper presents \graphalg{}, a domain-specific language that brings graph algorithms into databases as first-class citizens.
Our key insight is that linear algebra provides the ideal abstraction layer: matrix operations naturally express graph algorithms, compile to relational algebra (with aggregation), and enable aggressive optimization.
Unlike previous attempts, \graphalg{} satisfies four critical requirements:
(1)~\textbf{Expressive}: users can implement arbitrary algorithms by composing matrix operations;
(2)~\textbf{User-friendly}: algorithms require significantly less code than SQL or Pregel alternatives;
(3)~\textbf{Fully integrated}: algorithms embed directly in queries without data movement;
(4)~\textbf{Optimizable}: programs compile to query plans that leverage existing database optimizations.

We make three primary contributions:

\textbf{Language Design.}
We develop \graphalg{}, a language based on linear algebra with formal foundations in MATLANG~\cite{brijder_expressive_2019}.
\graphalg{} provides high-level matrix operations that compose naturally while compiling to efficient (extended) relational algebra (\cref{sec:lang}).

\textbf{Database Integration.}
We implement \graphalg{} in the AvantGraph~\cite{van_leeuwen_avantgraph_2022} query engine, where algorithms and queries share a unified representation.
This enables cross-boundary optimizations impossible in systems with separate algorithm pipelines (\cref{sec:integration}).

\textbf{Novel Optimizations.}
We introduce critical optimizations for execution of \graphalg{} programs: sparsity analysis maintains compact representations, loop-invariant code motion eliminates redundant computation, and in-place aggregation avoids materializing intermediate results (\cref{sec:opts}).

Our evaluation demonstrates that \graphalg{} significantly reduces code complexity compared to existing approaches while achieving superior performance on standard benchmarks.
On LDBC Graphalytics~\cite{iosup_ldbc_2016}, \graphalg{}/AvantGraph outperforms DuckDB and Neo4j on PageRank, SSSP, and WCC algorithms (\cref{sec:eval}).
The unified query-algorithm representation also enables powerful cross-boundary optimizations, allowing preprocessing steps to be pushed into algorithm execution without modifying the algorithm itself.

With \graphalg{}, graph databases finally get algorithm support \emph{done right}: uniting the expressiveness users demand, the performance they expect, and the seamless integration they require.
Graph algorithms are no longer second-class citizens forced to live outside the database, they are now where they belong, as first-class operations over graph data.

\section{Reviewing the Design Space}
Before detailing \graphalg{}'s design, we examine why existing approaches fail and what design choices enable our solution.
We review current database solutions (\cref{sec:related-analytics-in-db}), analyze computational models for graph algorithms (\cref{sec:other-languages}), and explain how \graphalg{} navigates this design space (\cref{sec:our-approach}).

\subsection{Limitations of Current Data Management Systems}\label{sec:related-analytics-in-db}
The options for graph analytics in existing databases depend on the specific system used.
An overview of existing approaches and their limitations is given in \cref{tab:existing-approaches}.
We describe each approach in more detail below.

\textbf{Algorithms library.}
Neo4j~\cite{neo4j_inc_graph_nodate} and ArangoDB~\cite{arangodb_inc_graph_nodate} bundle a library of commonly used graph algorithms with their systems.
Such a library is by far the easiest to use and typically offers good performance.
A major downside is that it can only be used if the exact algorithm required is included in the library.
To take the Neo4J Graph Data Science Library~\cite{neo4j_inc_graph_nodate} as an example: Its PageRank implementation has more than 10 parameters to select different variations of the algorithm, yet it does not allow redistribution of scores from sinks, which is trivial to do in a custom implementation.

\textbf{Pregel API.}
To accommodate the needs of users for which the algorithms library is not sufficient,
Neo4j~\cite{neo4j_inc_graph_nodate} also provides a Java API for writing algorithm in a vertex-centric style inspired by Pregel~\cite{malewicz_pregel_2010}.
The Pregel API is much more flexible, but as we show in \cref{sec:eval}, this comes with performance issues.
It appears that this is a fundamental problem to the vertex-centric model~\cite{khan_vertex-centric_2017,mcsherry_scalability_2015} and not merely an issue with the Neo4j implementation we have benchmarked.
Furthermore, because the body of a Pregel implementation can contain arbitrary Java code (including external libraries), applying high-level optimizations to the implementation is extremely challenging.
To the best of our knowledge, Neo4j does not attempt to do this.

\textbf{User-defined Operators.}
The Umbra database system supports user-defined operators written in C++~\cite{sichert_user-defined_2022}.
It has excellent performance, but the lack of isolation means a buggy implementation may crash or silently corrupt arbitrary memory locations.
Since the user can provide arbitrary C++ code, high-level optimization is out of the question.

\textbf{Recursive CTEs.}
Recursive common table expressions (CTEs) introduce recursion to SQL and thus greatly enhance its expressive power.
Because of their wide support in relational database systems, they are an obvious candidate for expressing graph algorithms.
Unfortunately, recursive CTEs are notoriously difficult to write~\cite{duta_another_2022}.
In practice, they are typically avoided in favor of external tools that are easier to program~\cite{lambrecht_trampoline-style_2025}.
In creating our benchmarking setup (see \cref{sec:eval}) we experimented with recursive CTEs in DuckDB.
We found that the algorithms were challenging to implement in this form and performed poorly.
Issuing multiple individual queries and buffering their results in temporary tables offered better performance in all cases.

Recent work has extended SQL with additional syntax to express complex control flow~\cite{lambrecht_trampoline-style_2025} and upset semantics~\cite{hirn_fix_2023} in recursive CTEs.
In future work we hope to explore whether recursive CTEs with this extended syntax are a viable compilation target for \graphalg{}.
This would allow running \graphalg{} programs on databases without explicit support, so long as they implement the necessary SQL extensions.

\textbf{Procedural Extensions to SQL.}
Several relational database systems, including PostgreSQL, offer a procedural language with support for embedded SQL queries.
This offers great flexibility and in theory would also allow for high-level analysis and optimization.
Unfortunately, it appears that implementations tend to focus more on functionality than performance.
Languages like pl/pgSQL are typically executed by simple interpreters, with the embedded SQL queries optimized and executed separately~\cite{gupta_procedural_2021}.
Some work even suggests compiling pl/pgSQL into plain SQL to avoid the overhead of the interpreter and context switching~\cite{duta_compiling_2019}.

\textbf{Algorithm DSL.}\label{sec:why-dsl}
Domain-specific languages (DSL) offer a compelling approach to algorithm support in databases.
DSLs provide dedicated syntax for common graph operations, making it easy to express algorithms.
Users can write arbitrary algorithms, though this may be constrained by the expressive power of the specific DSL. 
Similarly, performance and optimization opportunities depend on the design of the language, and on the execution strategy.
We are aware of two systems that implement this approach, TigerGraph~\cite{deutsch_tigergraph_2019} and Oracle PGX~\cite{boukham_spoofax_2023}.
Both systems are, sadly, proprietary: TigerGraph has only recently introduced a free Community Edition of its system. Oracle does not offer a free version at all.
Source code is not available, and the license agreements of both systems explicitly forbid running benchmarks.
As a result, public documentation on the internals of these systems is limited.
TigerGraph supports the GSQL language, a graph query language with support for iteration and accumulators~\cite{deutsch_aggregation_2020}.
Since algorithms can be expressed in GSQL, we assume that they are executed by the same query engine as regular queries.
However, based on the documentation\footnote{\url{https://docs.tigergraph.com/gsql-ref/4.2/querying/query-optimizer/}}, it appears that query optimization is limited.
For Oracle PGX specifically it is known that the pipeline for algorithms is separated from the query engine, so they are optimized differently than queries.

\subsection{Graph Algorithm Languages}\label{sec:other-languages}
Various languages for writing graph algorithms have been proposed in the literature.
These broadly fall into one of four categories based on the computational model, which we discuss below.

\textbf{General-Purpose}
languages such as Python or C++ are arguably the most popular approach today.
These languages have the clear advantage that they enjoy a large user base familiar with the syntax.
They also typically offer many libraries to simplify common graph analytics tasks.
The downside is that as a result of being very general, they are highly expressive and very challenging to analyze and optimize.
Our goal with \graphalg{} is to compile it to an efficient query plan, and such a transformation would only be possible for a tiny subset of valid programs.

\textbf{Vertex-Centric.}
Pioneered by Pregel~\cite{malewicz_pregel_2010}, the vertex-centric model is much more restrictive, especially in the way it accesses memory.
Fregel~\cite{iwasaki_fregel_2022} is an example of a language based on the TLAV (`Think Like A Vertex') paradigm.
Computation happens on a per-vertex level, and all communication between vertices happens via message passing.
This makes it eminently suitable for use in distributed settings, but on a single-node system this limitation can negatively impact performance.
We also observe an impedance mismatch between vertex-centric processing and relational algebra:
Where in the vertex-centric model each vertex acts autonomously, relational algebra applies the same operation over an entire relation.
This makes it harder to convert vertex-centric programs into relational algebra, and inhibits data-parallel processing in general.

\begin{figure}
  \includegraphics[width=\linewidth]{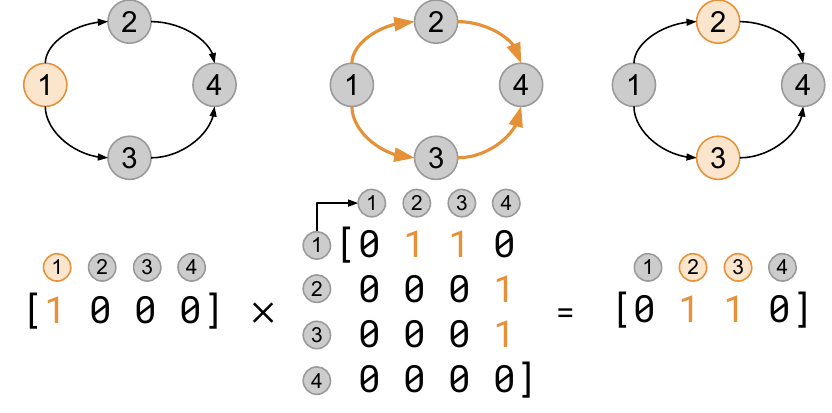}
  \caption{The connection between matrix multiplication and graph traversal.}
  \label{fig:matmul-graph-traversal}
\end{figure}

\textbf{Linear Algebra} is an established model for graph analytics with a rich literature~\cite{kepner_graph_2011}.
Linear algebra operations have a number of desirable properties.
Firstly, linear algebra operations are \emph{high-level} and \emph{composable}.
Because they are defined over entire matrices, query optimization opportunities are exposed and, in particular, execution is easily parallelized.
Various linear algebra operations have a straightforward translation to relational algebra~\cite{brijder_matrices_2022}.
Finally, many graph analytics tasks can be phrased as linear algebra computations.
For example, \cref{fig:matmul-graph-traversal} illustrates that matrix multiplication and graph traversal are closely related.
For these reasons, we adopt the linear algebra model for \graphalg{}.
The \matlang{}~\cite{brijder_expressive_2019} formal language for matrix manipulation has previously studied the linear algebra model, but it is purely a formal language without an implementation.
We discuss the relation between \graphalg{} and \matlang{} in \cref{sec:matlang}.
On the implementation side there is SuiteSparse:GraphBLAS~\cite{davis_algorithm_2019}, a suite of sparse matrix algorithms that can be composed to implement graph algorithms~\cite{mattson_lagraph_2019}.
SuiteSparse:GraphBLAS, however, is designed around its own internal data representations and runtime, so it requires a data wrangling step like other external tools.
It also requires all data to be loaded into main memory, and therefore cannot scale to graphs larger than main memory.

\textbf{Vertex and Edge Sets.}
This model is similar to the standard imperative model, but forces all graph operations to be done in bulk.
This limits expressivity enough that it becomes possible to perform high-level analysis and optimization.
Examples include Greenmarl~\cite{hong_green-marl_2012} and GraphIt~\cite{zhang_graphit_2018}.
We find languages based on the vertex and edge sets model interesting because they show excellent performance on a wide range of algorithms and workloads~\cite{azad_evaluation_2020}.
They can also be effectively analysed and optimized.
In many ways, the vertex and edge sets model resembles the linear algebra model:
A vertex set is similar to a (sparse) vector, as is an edge set to a matrix.
Linear algebra operations can be implemented in a language following the vertex and edge sets model.
The opposite relation, however, does not typically hold.
To take GraphIt~\cite{zhang_graphit_2018} as an example:
It supports the creation of arrays and writing to individual positions of that array in a parallel loop.
While mutable arrays can be very efficient in a shared-memory system, they do not map well to relational algebra.

\subsection{Our Approach}\label{sec:our-approach}
We argue that the right approach to algorithm support in databases is an integrated domain-specific language (DSL).
By building on the linear algebra computational model, \graphalg{} embodies our four principles for algorithm support \emph{done right}:

\textbf{Expressive and Optimizable.} 
\graphalg{} strikes a careful balance between expressivity and optimizability.
The user can encode arbitrary algorithms by composing high-level linear algebra primitives.
At the same time, \graphalg{} avoids operations that are difficult to optimize or not data-parallel, such as explicit mutation.
Linear algebra operations process full matrices, so they are easily parallelized.
Furthermore, a simplified core language (\cref{sec:core-lang}) and solid theoretical foundations (\cref{sec:matlang}) make \graphalg{} highly amenable to analysis and optimization.

\textbf{User-friendly.} \graphalg{} is designed specifically for writing graph algorithms.
The linear algebra operations have well-known semantics and can concisely represent graph algorithms.

\textbf{Fully Integrated.} 
\graphalg{} expressions have a straightforward translation into (an extension of) relational algebra, making it easy to integrate into an existing query processing pipeline.

\section{The \graphalg{} Language}\label{sec:lang}
In the next section we introduce the \graphalg{} language by explaining key language concepts based on example programs.
After this presentation of the `user' side of the language, we discuss the \emph{Core} language and its connection to \matlang{}~\cite{brijder_expressive_2019} that forms the formal basis of \graphalg{}.

\subsection{A Tour of \graphalg{}}
\begin{figure}
  \lstset{language=GraphAlg,frame=single,numbers=left,xleftmargin=2.5em}
\begin{lstlisting}
func Reach(G: Matrix<s, s, bool>,
           source: Vector<s, bool>) 
    -> Vector<s, bool> {
  v = source;
  for i in G.nrows {
    v += v * G;
  }
  return v;
}
\end{lstlisting}
  \caption{
    Reachability algorithm implemented in \graphalg{}.
  }
  \label{fig:reach}
\end{figure}

\begin{figure}
  \lstset{language=GraphAlg,frame=single,numbers=left,xleftmargin=2.5em}
\begin{lstlisting}
func SSSP(G: Matrix<s, s, trop_real>,
          source: Vector<s, trop_real>) 
    -> Vector<s, trop_real> {
  v = source;
  for i in G.nrows {
    v += v * G;
  }
  return v;
}
\end{lstlisting}
  \caption{
    Single source shortest path algorithm implemented in \graphalg{}.
  }
  \label{fig:sssp}
\end{figure}

The first example program is shown in \cref{fig:reach}.
Starting from a set of source vertices, it traverses the graph to find all vertices that are reachable from one of the source vertices through any number of hops.
As previously illustrated in \cref{fig:matmul-graph-traversal}, matrix multiplication is closely related to graph traversal.
If we take the vector \texttt{v} containing value `1' for all vertices that have been reached in previous iterations, and multiply it with the adjacency matrix that represents the graph, then the result is a vector with a `1' for every vertex that can be reached with a single hop.
The reachability algorithm executes the matrix multiplication in a loop and accumulates the results after every step.
The loop runs for as many iterations as there are vertices in the graph, and thus represents a full traversal of the graph.
Loops in \graphalg{} are always bounded, either by the number of vertices in the graph, or by a constant.
Recursion is forbidden to guarantee that all \graphalg{} programs are terminating.

The \texttt{s} in \texttt{Vector<s, bool>} is a \emph{dimension symbol}.
It represents the number of elements in the vector.
The use of abstract symbols allows for type checking of \graphalg{} programs without having to give concrete inputs to all function parameters.
Matrix multiplication, for example, requires that the number of columns on the left-hand side matches the number of rows on the right-hand side.
The \graphalg{} compiler can evaluate these requirements at compile time based on the dimension symbols, avoiding the need for runtime error detection.
The second type parameter (in this case \texttt{bool}) indicates a \emph{semiring}~\cite{kepner_graph_2011}.
The semiring defines the addition and multiplication operators for entries of the matrix.
For a boolean matrix these are $\lor$ and $\land$, respectively.
\graphalg{} also supports \texttt{int} for integers and \texttt{real} for real/floating-point numbers, with the usual operators.
Semirings are a powerful tool for defining graph algorithms.
Take for example the \emph{tropical semiring}, which instead uses the operators $\min(a,b)$ for addition and $+$ for multiplication.
If we modify the reachability algorithm of \cref{fig:reach} to use this semiring, we obtain the single source shortest path algorithm shown in \cref{fig:sssp}.
Note that because the tropical semiring defines the addition operator as $\min(a,b)$, the statement \texttt{v += v * G} should be understood to mean $v = min(v, v \cdot G)$.

\begin{figure}
  \lstset{language=GraphAlg,frame=single,numbers=left,xleftmargin=2.5em}
\begin{lstlisting}
func withDamping(degree:int, damp:real) -> real {
  return cast<real>(degree) / damp;
}
func PageRank(G: Matrix<s, s, bool>,
              damping: real,
              iterations: int) 
    -> Vector<s, real> {
  n = G.nrows;
  teleport = (real(1.0)-damping) / cast<real>(n);
  // Out degree with damping
  d_out = reduceRows(cast<int>(G));
  d = apply(withDamping, d_out, damping);
  // Vertices without outgoing edges
  connected = reduceRows(G);
  sinks = Vector<bool>(n);
  sinks<!connected>[:] = bool(true);

  pr = Vector<real>(n);
  pr[:] = real(1.0) / cast<real>(n);
  for i in int(0):iterations {
    // Score redistributed from sinks
    sink_pr = Vector<real>(n);
    sink_pr<sinks> = pr;
    redist = (damping / cast<real>(n)) 
      * reduce(sink_pr);
    // New PageRank score
    w = pr (./) d;
    pr[:] = teleport + redist;
    pr += cast<real>(G).T * w;
  }
  return pr;
}
\end{lstlisting}
  \caption{
    PageRank implemented in \graphalg{}.
  }
  \label{fig:pr}
\end{figure}

In \cref{fig:pr} we give an implementation of PageRank~\cite{pagerank}.
We highlight a few additional language constructs used in the algorithm:
\begin{itemize}[leftmargin=*]
  \item \texttt{cast<T>} (line 2, 9, etc.) casts between different semirings, such as integer to real conversion.
  \item \texttt{reduceRows} (line 11, 14) collapses a matrix to a column vector, summing elements using the addition operator.
  \item \texttt{apply(f, M, c)} (line 12) applies a scalar function \texttt{f} to each element of \texttt{M}, where
        \texttt{c} is an additional scalar value passed as the second argument to \texttt{f}.
        The output matrix is defined as $O_{ij} = f(M_{ij}, c)$.
  \item \texttt{M[:] = c} (line 16, 19, 28) replaces every entry of \texttt{M} by scalar \texttt{c}.
  \item \texttt{A<M> = B} (line 16, 23) assigns elements from \texttt{B} to the same position in \texttt{A} iff \texttt{M} has a nonzero value at that position.
  \item \texttt{reduce} (line 25) sums all elements of a matrix to a scalar.
  \item \texttt{(.f)}  (line 27) represents pointwise function application.
  \item \texttt{M.T} (line 29) is the transpose of \texttt{M}.
\end{itemize}

The PageRank implementation demonstrates that \graphalg{} has the expressive power to encode non-trivial algorithms by combining linear algebra operations.
The imperative programming paradigm coupled with high-level operations such as masking and matrix-to-vector reduction and a high-level data model allow for a clear and concise definition of the algorithm.

\subsection{Core Language}\label{sec:core-lang}
The \graphalg{} compiler simplifies \graphalg{} programs to a subset of \graphalg{}, which we denote by \graphalg{} \emph{Core}.
\graphalg{} Core is used for analysis, optimization and translation into relational algebra.
The Core language is defined by the grammar:

\begin{align*}
  E ::= & \ M                                  & \text{(matrix variable)}           \\
  |     & \ E^{*}                              & \text{(transpose)}                 \\
  |     & \ \text{diag}(E)                     & \text{(diagonalizate a vector)}    \\
  |     & \ \text{apply}[f](E_1,\ \cdots,E_n)  & \text{(pointwise application)}     \\
  |     & \ E_1 \cdot E_2                      & \text{(matrix multiplication)}     \\
  |     & \ \mathbf{1}(r, s)                   & \text{(one-vector)}                \\
  |     & \ \text{for}[b](s, E_1,\ \cdots,E_n) & \text{(loop over dimension)}       \\
  |     & \ \text{for}[b](l, E_1,\ \cdots,E_n) & \text{(loop over range)}           \\
  |     & \ \text{pickAny}(E)                  & \text{(keep first nonzero in row)} \\
  b ::= & \{ M_1 = E_1,\ \cdots, M_n = E_n \}  & \text{(loop body)}                 \\
  f ::= & (C_1,\ \cdots,C_n) \ e               & \text{(pointwise function)}        \\
  e ::= & \ C                                  & \text{(scalar variable)}           \\
  % CastDimOp (covered by literal case)
  |     & \ r(l)                               & \text{(scalar literal)}            \\
  |     & \ e_1 \ \{+,\cdot,-,/,=\} \ e_2      & \text{(scalar arithmetic)}         \\
  |     & \ \text{cast}(r, e)                  & \text{(scalar cast)}               \\
\end{align*}

Most of the operations are standard linear algebra concepts such as matrix multiplication or transpose, following the usual semantics~\cite{lang_introduction_1986}.
We describe the behavior of the non-standard operations in more detail below.

$\mathbf{1}(r, s)$ constructs a vector with dimension $s$ and semiring $r$.
The elements of the vector are initialized with the multiplicative identity, which is `1' in most semirings.

The `for' operation defines a loop with one or more state variables $M_1,\ \cdots,M_n$.
The loop body consists of one expression per state variable that defines how to produce the value for the next iteration of the loop from the previous state.
Expressions can refer to the current loop state by referencing one of the state variables.
The initial values for the state variables are given by $E_1,\ \cdots,E_n$ in the loop definition.
The number of iterations that the loop will run for is defined either by a matrix dimension $s$ or an integer literal $l$.
The result of the loop expression is the final value of the first state variable $M_1$.

The `pickAny' operation takes an input matrix and zeros out all but one element per row (if the row has at least one nonzero element).
This primitive is useful for \emph{leader selection} problems, such as connected components algorithms that need to choose one representative vertex per component.
An example use of the `pickAny' operation is presented in line 8 of \cref{fig:cypher-embed}.

More details on \graphalg{} Core, such as the conversion from the full \graphalg{} language or the transformation to relational algebra can be found in the language specification~\cite{de_graaf_graphalg_nodate}.

\subsection{Theoretical Foundations}\label{sec:matlang}
\graphalg{} Core is based on \matlang{}~\cite{brijder_expressive_2019} and its extension for-\matlang{}~\cite{geerts_expressive_2021}, which are formal languages for matrix manipulation.
They are positioned as a matrix counterpart to the relational algebra.
\matlang{} is to \graphalg{} what relational algebra is to SQL:
\matlang{} provides a principled theoretical foundation for \graphalg{}.
This theory provides a translation of \graphalg{} programs into (an extension of) relational algebra~\cite{brijder_matrices_2022}.
\graphalg{} in turn can be seen as the \emph{first implementation} of \matlang{}, which previously existed only as a language for theoretical analysis.
Most of the operations follow the definition from \matlang{}~\cite{brijder_expressive_2019} with only minor changes.

\graphalg{} makes three key design decisions that distinguish it from \matlang{} and enable efficient database integration.

\textbf{Optimizable loop decomposition.}
for-\matlang{}~\cite{geerts_expressive_2021} provides a highly expressive loop construct.
\graphalg{} decomposes this further into two orthogonal primitives: (1) a bounded loop as in~\cite{chandra_programming_1981}, and (2) the dedicated `pickAny' operation for leader selection.
This design is motivated by compilation to relational algebra and query optimization:
`pickAny' parallelizes trivially across all rows and compiles to standard aggregation, making it amenable to existing database optimizations.
The separation enables the query optimizer to reason about and optimize each primitive independently, something that would be impossible with for-\matlang{}'s more complex loop semantics.
This decomposition maintains the same expressive power as for-\matlang{}~\cite{de_graaf_graphalg_nodate}.

\textbf{Simultaneous induction for algorithm efficiency.}
Unlike for-\matlang{}, \graphalg{} loops support simultaneous induction~\cite{ebbinghaus_mathematical_2021}, carrying multiple state variables across iterations.
This design choice, adopted from~\cite{chandra_programming_1981}, is essential for expressing graph algorithms naturally and efficiently.
Consider breadth-first search: it must track both visited nodes and the frontier to explore in the next iteration.
Without simultaneous induction, algorithms must encode multiple values into a single matrix using complex workarounds, significantly complicating both implementation and optimization.

\textbf{Practical semiring support with casting.}
Real-world graph analytics demand heterogeneous data types, yet \matlang{} assumes all matrices share the same semiring (typically $\mathbb{C}$).
\graphalg{} introduces a `cast' operation to support practical analytics workloads where integers, floating-point numbers, and booleans commonly intermix.
For example, computing unweighted shortest paths requires casting a boolean adjacency matrix to a tropical semiring.
This feature is critical for database integration, where typed data is fundamental.

\section{Integration with AvantGraph}\label{sec:integration}
We implement \graphalg{} support into the AvantGraph~\cite{van_leeuwen_avantgraph_2022} graph query engine to enable the execution of user-defined graph algorithms from Cypher directly over the stored graph.

\subsection{Embedding in Queries}
\begin{figure}
  \lstset{
    language=GraphAlgInCypher,
    moredelim=[is][\textcolor{pgreen}]{\%\%}{\%\%},
    moredelim=[is][\textcolor{pred}]{\%\%\%}{\%\%\%},
    frame=single,
    numbers=left,
    xleftmargin=2.5em
  }
\begin{lstlisting}
WITH ALGORITHM "
func WCC(G: Matrix<s, s, bool>) 
    -> Matrix<s, s, bool> {
  id = Vector<bool>(G.nrows);
  id[:] = bool(true);
  label = diag(id);
  for i in G.nrows {
    label = pickAny(label + G * label);
  }
  return label;
}" CALL WCC((:%%%Person%%%)-[:%%SEND_MSG%%]->(:%%%Person%%%))
RETURN row.name, col.name AS label
\end{lstlisting}
  \caption{
    Weakly connected components algorithm embedded in Cypher.
  }
  \label{fig:cypher-embed}
\end{figure}

Algorithm definitions can be embedded directly into Cypher queries.
We introduce a \texttt{WITH ALGORITHM} clause as an extension of Cypher that allows the user to define \graphalg{} programs inside queries.
The functions defined in the program can then be invoked from Cypher using the \texttt{CALL} syntax that is normally used for built-in procedures.
\cref{fig:cypher-embed} shows an example query with an embedded \graphalg{} program.
The graph is passed to the algorithm as an argument, which can include additional filtering such as label predicates.
While Cypher is restricted to node/relationship patterns as \texttt{CALL} arguments, internally AvantGraph supports arbitrary queries as arguments to \graphalg{} programs.

\subsection{Integration into Query Pipeline}

\begin{figure}
  \includegraphics[width=\linewidth]{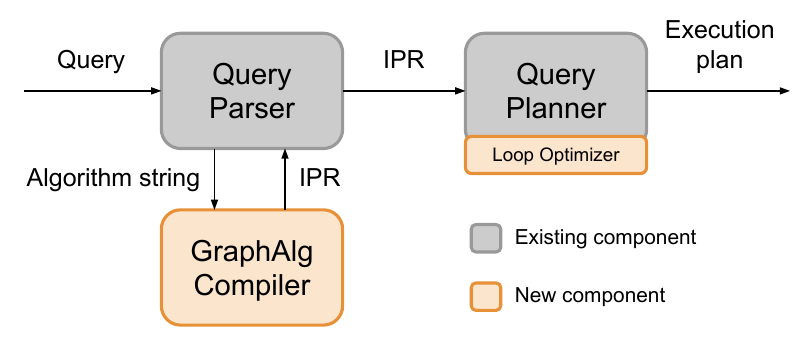}
  \caption{
    Integration of \graphalg{} compiler and loop optimizations within AvantGraph.
    Pipeline stages after the query planner are not affected and are omitted from the diagram.
  }
  \label{fig:integration}
\end{figure}

\cref{fig:integration} shows how the \graphalg{} compiler is connected to the AvantGraph query pipeline at a high level.
IPR stands for `Internal Plan Representation', the logical query plan representation used internally by AvantGraph.
The \graphalg{} compiler outputs IPR as a final step, but internally performs its analyses and optimizations on a \graphalg{} Core representation.
This design ensures a loose coupling between the \graphalg{} compiler and the AvantGraph query pipeline.
Some optimizations related to loops, described in detail in \cref{sec:licm} and \cref{sec:loop-agg}, must be applied to physical rather than logical plans.
Therefore, they are kept separate from the \graphalg{} compiler and implemented as an extension to the query planner.
Note that the \graphalg{} compiler on its own can already transform \graphalg{} programs into valid IPR: the loop optimizer only improves the quality of the generated plans.

\subsection{Unified Representation}\label{sec:unified-ir}
Another important take-away from \cref{fig:integration} is that after the \graphalg{} compiler is finished, query and algorithm are both represented in IPR.
At this point a single IPR query represents both the query and the embedded algorithm.
Queries and algorithms become indistinguishable from one another.
This has a number of important implications for the execution of algorithms.
Firstly, the existing query planner and runtime can be reused for algorithm execution.
Compared to other systems that have separate pipelines for query and algorithm processing, engineering effort to implement and maintain algorithm support is greatly reduced.
Additionally, there is no cost to exchange data between query and algorithm.
As shown previously in \cref{fig:cypher-embed}, large intermediates (even the full graph) can be passed from query to algorithm without having to materialize them first.
Because the query planner operates over a unified representation, it can even apply optimizations across the query/algorithm boundary, such as pushing query-side filters on input arguments down into algorithms.

\subsection{Changes to AvantGraph}\label{sec:impl-support}

\begin{figure}
  \centering
  \includegraphics[width=.5\linewidth]{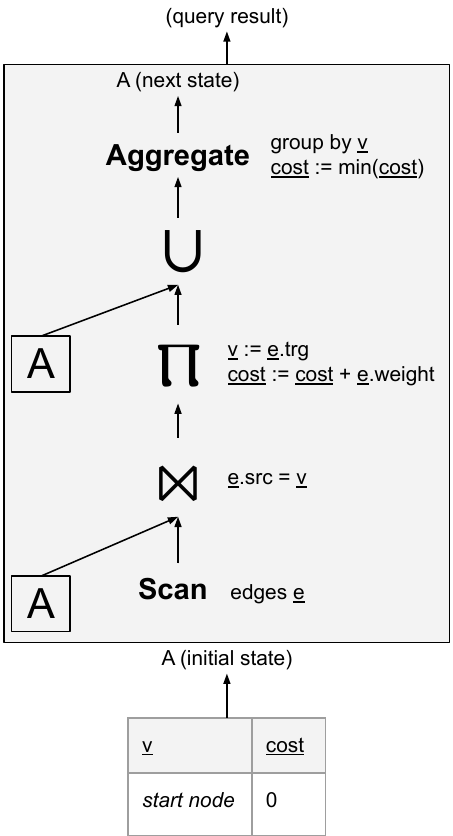}
  \caption{
    Logical query plan for a single source shortest path algorithm.
  }
  \label{fig:loop}
\end{figure}

Since most operations in \graphalg{} compile to standard database operations such as joins and aggregations, only minimal changes were needed to make AvantGraph a fully functional target for the \graphalg{} compiler.
The loop construct in \graphalg{}, a requirement for iterative algorithms, is the only operation that cannot be emulated.
To support it we extend the AvantGraph internal query representation with a `Loop' operator that is closely modeled after the \graphalg{} Core definition.
Inputs to the operator are the initial states of the loop variables, as well as the number of iterations that the loop will execute.
The loop operator embeds one subquery for every loop variable which computes the next value of that variable.
Together these subqueries encode the loop body of the original \graphalg{} program.
In \cref{fig:loop} we show an example of using a loop operator in a logical query plan expressing the single-source shortest path algorithm using otherwise standard query operators.
While a loop can have multiple state variables, it always has exactly one output, like in \graphalg{} Core.
The final output is the value of one of the state variables (state A in the example).
This is important because the AvantGraph query optimizer requires query plans to be trees.

\section{Query Optimization}\label{sec:opts}
We describe the key optimizations we perform on \graphalg{} programs to achieve competitive performance in AvantGraph.
As shown in \cref{fig:integration}, the \graphalg{} compiler performs the sparsity analysis, whereas the loop optimizations are implemented in the AvantGraph query planner.
These optimizations are in addition to the standard query optimization techniques such as predicate pushdown and join order optimizations that are already applied to algorithms through the unified intermediate representation (\cref{sec:unified-ir}).

\subsection{Sparsity}
The \graphalg{} compiler represents a matrix as a table with three attributes (row, column, value).
In the most naive conversion to relational algebra, the table contains a tuple for every valid combination of row and column.
In the context of AvantGraph, the dimensions of matrices correspond to the number of vertices $V$ in the graph.
This implies that a matrix table has $V^2$ entries, which is prohibitively expensive for large graphs.
Fortunately, the adjacency matrix of most real-world graphs is very sparse, and hence \graphalg{} programs often operate on matrices where most elements are zero.
To save both storage space and compute time, one typically stores only the non-zero elements, which we refer to as \emph{sparse matrix representation}.
\graphalg{} is designed such that most operations can operate directly over a sparse matrix representation, rather converting to an expensive dense representation first.

For a simple example, consider the \graphalg{} expression \texttt{A (.*) B}.
This expression can be converted into the following relational algebra expression, regardless of whether the underlying tables omit any zero elements: $\pi_{A_v \cdot B_v}(A \bowtie_{r, c} B)$.
If $A$ and $B$ are sparse, implicit zeros in the input result in implicit zeros at the same positions in the output.
The same holds for matrix multiplication over any semiring.
Not all operations can directly operate over the sparse representation though.
For example, the expression \texttt{A (.==) B;} has a non-zero value for position $(r, c)$ even if $A_{rc} = B_{rc} = 0$.

The \graphalg{} compiler implements an analysis to track for every matrix in the program whether it is potentially sparse (omitting some zero-value tuples).
By default, all matrices are kept in sparse representation.
If an operation is encountered that requires a dense matrix, the value is explicitly made dense (adding zero-value tuples).
This approach keeps data in sparse representation whenever possible, while still preserving the correct semantics for operations that require dense inputs.

\subsection{Loop-Invariant Code Motion}\label{sec:licm}
Loop-invariant code motion~\cite{aho_compilers_2007} is a classic program optimization technique.
The optimization finds an expression computed inside the loop that produces the same value at every iteration, and instead evaluates it once before the loop to avoid recomputing it multiple times.
We carry over this program optimization technique to query plans.
The challenge in this context is that intermediate results can be large tuple streams, which must be cached in a buffer to allow reading them repeatedly.
While finding loop-invariant query fragments inside loops is trivial, it is less trivial to judge if moving them will improve query performance.
Consider the query plan in \cref{fig:loop}.
The loop body contains an edge scan node that is evaluated repeatedly and is loop-invariant.
Still, pulling the edge scan outside the loop is unlikely to be beneficial:
the scan operator produces a stream of tuples, which must be cached in main memory.
Given a large enough graph, the cost of storing all tuples is prohibitively high.

Our modified implementation of loop-invariant code motion only extracts loop-invariant query fragments that produce a hash table.
This way we get the required caching for free, because the hash table can be probed repeatedly.
It also avoids introducing additional pipeline breakers to plans.
Because query fragments may themselves contain hash table build operators, we always extract the largest possible query fragments first.
Queries generated from \graphalg{} often make heavy use of (hash) aggregation, which makes our approach highly effective.
It is not optimal though, as there exist expensive query fragments that produce small intermediate results, and thus would be worth caching.

\subsection{Aggregation across Loop Iterations}\label{sec:loop-agg}
\begin{figure}
  \centering
  \includegraphics[width=.6\linewidth]{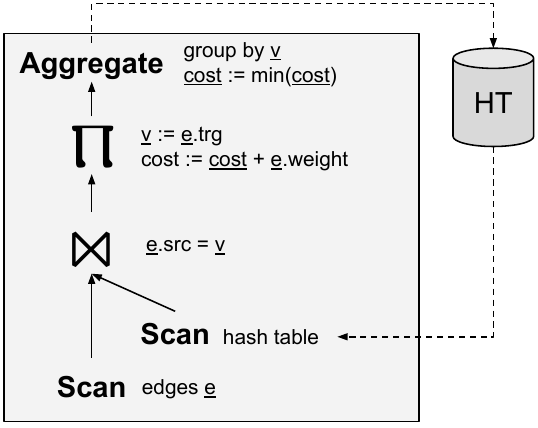}
  \caption{
    Loop body of single source shortest path algorithm with in-place aggregation.
  }
  \label{fig:in-place}
\end{figure}

A common pattern in many graph algorithms is the aggregation of values from multiple loop iterations.
The single source shortest path algorithm is one such example, as can be seen in \cref{fig:loop}, where the loop body has an `Aggregate' operator at the root.
This aggregation has two inputs, joined into one with a union operator: (1) on the left hand side, the new costs for vertices computed in the current iteration, and (2) the existing costs for vertices collected in previous iterations.
Every iteration, state A is read, combined with additional results, and then aggregated again.
Assuming that state A is already stored in a hash table, reading the table and aggregating into another hash table is unnecessary.
We implement a loop optimization pass to instead add the additional results directly to the existing hash table.
\cref{fig:in-place} shows the loop body for the single source shortest path algorithm after applying the in-place aggregation optimization.
Note that the hash table is populated with initial state before the loop starts.
It is also read after loop termination to obtain the final loop result.

Loops with in-place aggregation also become candidates for \emph{fixpoint analysis}.
At the end of each loop iteration, the AvantGraph query runtime checks if the state of any loop variables has been modified since the previous iteration.
If there were no changes to the loop state during an iteration, then the loop has reached a fixed point where there will be no further state changes, causing the runtime to terminate the loop.
For this reason, \graphalg{} programs can often safely define loops to run for as many iterations as there are vertices in the graph (see for example \cref{fig:sssp}), relying on the runtime to exit the loop early.

\section{Experiments}\label{sec:eval}
We compare our solution, \graphalg{} embedded in AvantGraph, to other database systems that support user-defined algorithms.
These systems include graph-native (Neo4j) as well as relational systems (PostgreSQL, DuckDB).
The algorithms and datasets used for the evaluation are from LDBC Graphalytics~\cite{iosup_ldbc_2016}, a graph analytics benchmark.
Graphalytics is designed for graph analytics platforms, not specifically databases, but it does provide a reference implementation for PostgreSQL-compatible database servers.
Our evaluation is based on five graph algorithms as specified in the benchmark:
\begin{itemize}
  \item \textbf{Breadth-first search (BFS)}: Per-vertex minimum number of hops from the source vertex.
  \item \textbf{PageRank (PR)}: Approximate vertex importance using PageRank~\cite{pagerank}, with redistribution from sinks.
  \item \textbf{Weakly connected components (WCC)}: Finds all weakly connected components in the graph, assigning each vertex in a component the same label.
  \item \textbf{Community Detection using Label Propagation (CDLP)}: Finds weakly connected communities in the graph~\cite{raghavan_near_2007}.
        The graphalytics specification~\cite{iosup_ldbc_2023} modifies the algorithm slightly to make it parallel and deterministic.
  \item \textbf{Single-source shortest paths (SSSP)}: Total cost of the weighted shortest path from a source vertex to all vertices.
\end{itemize}

We omit the sixth algorithm in Graphalytics, Local clustering coefficient (LCC).
While \graphalg{} can express LCC, it is also easily expressed in Cypher or non-recursive SQL, and therefore not an interesting algorithm to consider here.

Next to LDBC Graphalytics, we also run PageRank on the OpenAIRE dataset for energy planning~\cite{baglioni_skg-if_2025} with different preprocessing steps.
This shows how systems perform on a realistic graph that requires transformations such as filtering or data cleaning before running an algorithm.

We evaluate systems based on two metrics:
\textbf{(1) Code Complexity}: How easy is it to implement the algorithms from the LDBC Graphalytics benchmark?
\textbf{(2) Execution time}: We record query execution times of the algorithms on several datasets from Graphalytics, as well as the OpenAIRE dataset for energy planning.

All tested implementations build on general-purpose APIs that can be used to implement arbitrary algorithms.
We explicitly do not consider algorithms from Neo4j's Graph Data Science library~\cite{neo4j_inc_graph_nodate} or DuckPGQ's algorithms~\cite{wolde_duckpgq_2023}, as these are algorithm-specific solutions:
They can not be used to implement user-defined algorithms, and apply only if a user needs that exact algorithm.

\subsection{Systems Tested}
\subsubsection*{PostgreSQL}
We include PostgreSQL 15.2 because a reference implementation of LDBC Graphalytics is available\footnote{\url{https://github.com/ldbc/ldbc_graphalytics_platforms_umbra}}.
This implementation encodes the loop control flow in Java, issuing queries to the PostgreSQL server to traverse the graph and perform computation.
State kept between loop iterations is stored in temporary tables.
The Java component must be executed client-side, so this is not a self-contained solution that works entirely in PostgreSQL.
An alternative implementation could avoid this by implementing the algorithm in PL/pgSQL.
This has the additional advantages of (1) reducing the number of roundtrips between client and are server, and (2) avoids parsing queries every loop iteration.
For the graphs we test here, however, these costs are negligible compared to query execution time.

\subsubsection*{DuckDB}
DuckDB is not specifically a graph database, nor does it support procedural control flow.
It is, however, optimized for analytical processing, and has a user-friendly Python API.
While data exchange between Python and SQL is inconvenient, the combination of the language provide robust query performance and great expressivity.
Since there was no existing Graphalytics setup for DuckDB, we develop our own by adapting the reference PostgreSQL implementation.
We use DuckDB v1.4, the latest release at the time of the experiments.
We make two non-trivial changes in the DuckDB version compared to PostgreSQL.
Firstly, The postgreSQL implementation of Weakly-Connected Components uses a recursive CTE.
In DuckDB this fails with an out of memory error on all graphs we have tested.
We replace this with an iterative approach using temporary tables, which can handle all datasets we have tested.
Secondly, all temporary tables are kept in main memory to avoid slowing down queries due to disk I/O.

\subsubsection*{Neo4j}
Neo4j is a popular graph database with support for user-defined algorithms.
We build our implementation on the Pregel API~\cite{neo4j_inc_graph_nodate}, which allows defining custom algorithms in Java.
We use Neo4j Community Edition 2025.08 with Graph Data Science library 2.21, the latest release at the time of the experiments.
Our custom algorithms are compiled into a Neo4j plugin and are then invoked using the Cypher \texttt{CALL} syntax.
We highlight two issues with Neo4j that not only affect our experiments, but also have ramifications for real-world use.
Firstly, Neo4j does not currently support loading plugins on-demand.
To load in custom algorithms, the database server must be shut down and then restarted.
This makes iterating on algorithm design tedious, and all but impossible on multi-tenant instances.
Perhaps as a result, the Pregel API is not available in Neo4j's managed cloud offering.
Secondly, algorithms do not run on the graph storage used for regular queries.
Instead, the user creates a separate \emph{in-memory} graph object, and then runs the algorithm on this graph object.
Neo4j therefore requires that the graph fits entirely in main memory, severely impacting scalability.
We expect that in real-world scenarios, users will often need to unload graphs directly after executing an algorithm to reclaim memory for other tasks.

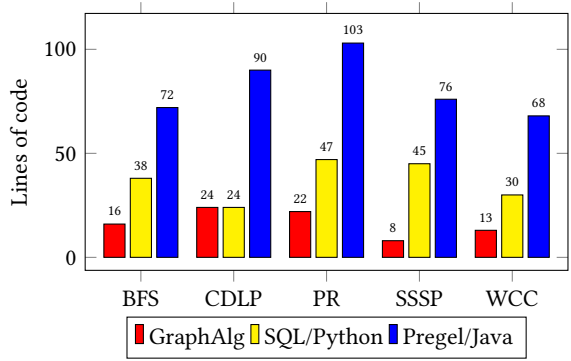
\begin{figure}
  \centering
  \input{loc-comparison.tex}
  \caption{
    Lines of code per Algorithm, excluding whitespace and comments.
    \graphalg{} requires the fewest lines of code on all algorithms, up to 9.5x fewer than alternative implementations.
  }
  \label{fig:loc-compare}
\end{figure}

\subsection{Experiment 1: \graphalg{} Reduces Code Complexity}

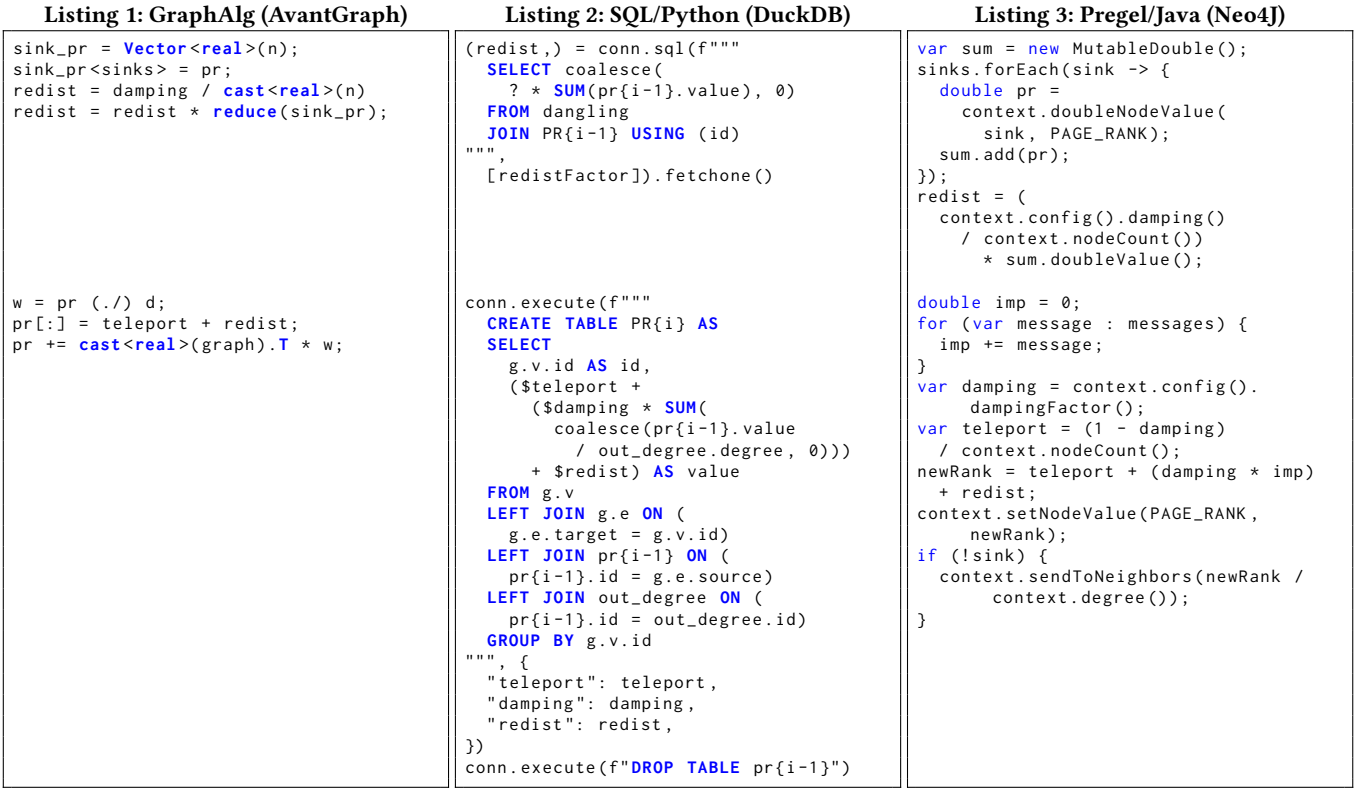
\begin{figure*}
  \input{pr-body.tex}
  \caption{
    Code complexity of different implementations of the PageRank loop body.
    The \graphalg{} version has vastly lower code complexity and is much shorter.
  }
  \label{fig:pr-body-compare}
\end{figure*}

To illustrate the differences in code complexity between the various systems, we present a side-by-side comparison of the loop body for the PageRank algorithm in \cref{fig:pr-body-compare}.
All three implementations compute the updated PageRank score in the same way.
First, the score to redistribute from sinks is calculated.
Then, the new score is computed by summing the teleportation and redistribution factors and adding the scores of adjacent vertices.

\graphalg{} stands out in terms of code size, with only 7 lines of code to perform both operations.
The SQL version is complicated by the Python code that orchestrates it.
Furthermore, a complex multi-way join is needed to bring together all data needed to compute the updated PageRank score.
The Pregel implementation consists of imperative Java code that is run in parallel for multiple vertices.
Its complexity follows mainly from the explicit exchange of messages between vertices, and the explicit iteration over sinks and messages.

A similar pattern repeats across the various algorithms.
In \cref{fig:loc-compare} we show the number of lines of code needed to express each algorithm in the various systems.
We omit the SQL/Java case for the PostgreSQL implementation since it is very similar to the DuckDB version.
Whitespace and comments are omitted from the count, as are the Gradle build files required to build the Neo4j plugin.
We observe that \graphalg{} programs are the smallest across the board.
The Pregel/Java implementations are the most verbose, owing to the general verbosity of the Java language.
For the CDLP algorithm, \graphalg{} and SQL/Python curiously end up with the exact same lines of code, yet the respective implementations are very different.

The results above support our hypothesis that algorithms are easier to implement in \graphalg{} than SQL/Python or Pregel/Java.
Further in-depth insights could be drawn from a user-study on programming in \graphalg{}.
This is left as future work.

\begin{table*}
  \small
  \centering
  \input{table.tex}
  \caption{
    Processing times on LDBC Graphalytics datasets (max size S) for \graphalg{}/AvantGraph (GAA), DuckDB (DDB) and Neo4j (N4J).
    Lowest execution time per dataset/algorithm in bold, second best underlined.
    \graphalg{}/AvantGraph dominates on the PR, SSSP and WCC algorithms, and is competitive on BFS.
  }
  \label{tab:proc-time}
\end{table*}

\subsection{Experiment 2: \graphalg{}/AvantGraph is Competitive on LDBC Graphalytics}

\begin{figure*}[t!]
  \centering
  \begin{minipage}{0.49\textwidth}
    \centering
    \input{execution_times_filter.tex}
    \caption{
      Execution time of PageRank on citations in OpenAIRE dataset for energy planning~\cite{baglioni_skg-if_2025} with self-edges removed.
      \graphalg{}/AvantGraph scales better than DuckDB while avoiding the preprocessing overhead of Neo4j.
    }
    \label{fig:cross-opt-filter}
  \end{minipage}
  \hfill
  \begin{minipage}{0.49\textwidth}
    \centering
    \input{execution_times_aggregate.tex}
    \caption{
      Execution time of PageRank on citations in OpenAIRE dataset for energy planning~\cite{baglioni_skg-if_2025} with duplicate edges removed.
      \graphalg{}/AvantGraph scales better than DuckDB and is faster than Neo4j up to a high number of iterations.
    }
    \label{fig:cross-opt-aggregate}
  \end{minipage}
\end{figure*}

We measure the execution time of the BFS, CDLP, PR, SSSP and WCC algorithms on all datasets from LDBC Graphalytics up through size `S'.
The results are collected using the LDBC Graphalytics benchmark driver.
We record the \emph{processing time}: the time required to execute the actual algorithm, excluding time to import the graphs into the system or (in the case of Neo4j) load them into main memory.
Experiments were run on a laptop with an Intel i7-11800H CPU with 8 cores and 16 hyper-threads, 32GB main memory and an NVMe SSD.
Results are shown in \cref{tab:proc-time}.

\subsubsection*{PostgreSQL}
PostgreSQL is consistently and significantly slower than the other systems.
Results are omitted from the table due to space constraints.
We refer interested readers to our benchmark setup to replicate these results.

\subsubsection*{OOM errors in Neo4j}
CDLP, SSSP and WCC in Neo4j fail with ``out of memory'' errors on datasets with many vertices (13M).
This suggests that the Pregel interface has a high per-vertex overhead.
Additionally, Neo4j requires that the full graph is loaded into main memory, leaving less space for intermediate state of algorithms.

\subsubsection*{BFS}
Performance of \graphalg{}/AvantGraph and DuckDB is comparable, with Neo4j a distant third.
On the larger datagen-7\_* datasets \graphalg{}/AvantGraph falls behind DuckDB.
Edge lookup is the dominant bottleneck for \graphalg{}/AvantGraph, likely as a result of a less efficient disk format compared to DuckDB.
For the datagen-7\_8-zf dataset, the graph is more than 5 times as large when loaded in AvantGraph.

\subsubsection*{CDLP}
For this algorith both DuckDB and Neo4j perform much better than \graphalg{}/AvantGraph.
The query plans for DuckDB generated from SQL are very similar to the AvantGraph query plans generated from \graphalg{}, so the difference in performance cannot be explained by a poor translation.
Instead, we observe that \graphalg{}/AvantGraph is substantially slower than DuckDB when computing a large hash aggregation.
This operation is repeated for every loop iteration, and therefore has a major influence on processing time.
This highlights an important performance issue in AvantGraph to be addressed.

\subsubsection*{PR}
\graphalg{}/AvantGraph shows the best performance on this benchmark, although DuckDB is not far behind.
The dota-league dataset is an exception, where DuckDB is faster than \graphalg{}/\allowbreak{}AvantGraph.
We attribute this to the extremely low vertex count (61K) compared to the number of edges (50M).
In AvantGraph this leads to contention between different threads performing hash aggregation.
DuckDB appears to handle this better.

\subsubsection*{SSSP}
\graphalg{}/AvantGraph is faster than both DuckDB and Neo4j, similar to PR.
A key advantage of \graphalg{}/AvantGraph over DuckDB is that it maintains a single result table populated over multiple iterations of the loop, whereas the DuckDB version must construct a fresh table every iteration.

\subsubsection*{WCC}
\graphalg{}/AvantGraph dominates the benchmark.
The difference between \graphalg{}/AvantGraph and DuckDB is larger because the algorithm is little more than repeated aggregation until a fix-point is reached.
It shows even more clearly than SSSP how important in-place aggregation is to efficient execution.

\subsubsection*{In Summary}
\graphalg{}/AvantGraph outperforms the other systems on the PR, SSSP and WCC algorithms and shows competitive performance for BFS.
Future work is needed to improve hash aggregation performance in AvantGraph so that it can support efficient execution of CDLP.

\subsection{Experiment 3: \graphalg{}/AvantGraph Robustly Handles Preprocessing}

Real graph analysis requires flexible preprocessing, but traditional systems force an awkward choice: either materialize cleaned data (expensive for exploratory workflows) or hardcode preprocessing into algorithms (inflexible).
Such preprocessing is ubiquitous in practice—filtering edges by time windows, removing bots or test accounts, handling duplicates and self-loops, normalizing weights, or extracting domain-specific subgraphs.
\graphalg{}/AvantGraph's unified IR enables a third option: query-side preprocessing that the optimizer can fuse with algorithm execution.
This experiment evaluates whether this integration delivers practical benefits.

We observed the importance of flexible preprocessing in the SciLake\footnote{\url{https://scilake.eu}} project, where PageRank is run on citation graphs to find high-impact publications.
The graphs have data quality issues requiring cleanup, but the right cleaning procedure was discovered through trial-and-error.
This iterative exploration demands query-side preprocessing without modifying algorithm definitions.

We hypothesize that \graphalg{}/AvantGraph's cross-optimization (\cref{sec:unified-ir}, \cref{sec:licm}) will enable query-side preprocessing with minimal overhead compared to systems that either recompute preprocessing per iteration (DuckDB) or require expensive materialization (Neo4j).

\subsubsection*{Experimental Setup}
We select the largest citation graph used in the SciLake project, the OpenAIRE dataset for energy planning~\cite{baglioni_skg-if_2025}, a knowledge graph of scientific publications in the energy planning field.
We consider two use cases that require preprocessing of the graph:

\textbf{No self references.}
We reduce the influence of authors that cite their own papers by removing citations where the first author of the citing and the cited papers match.
This is exposed as a boolean flag per edge. The vast majority of citations are \emph{not} self-referential (99\%), making this a simple low-selectivity predicate.

\textbf{No duplicate edges.}
Multiple citation edges between two papers indicates a data cleaning issue.
This is addressed by removing duplicate edges from the graph.
It is a significantly more expensive filter as it requires an aggregation over all edges in the graph.

These two cases test different optimization capabilities: (1) filter pushdown for cheap predicates, and (2) loop-invariant code motion for expensive aggregations.

With \graphalg{}/AvantGraph we can specify the required preprocessing on the query side as an argument to the algorithm, without making changes to the underlying algorithm.
We rely on the query optimizer to pick a query plan that efficiently incorporates the preprocessing step with algorithm execution.
For DuckDB we modify the PageRank implementation used in the previous experiment to take an \emph{edge view} as an additional parameter, and to use this instead of directly scanning the edge table.
This way the query optimizer can also take the preprocessing into account.
For Neo4j we can use the PageRank implementation as-is, and instead do all preprocessing when loading nodes and relationships from disk into main memory.
Preprocessing is thus a separate step that must be completed before the algorithm can be executed. 
Preprocessing time is included in the measured \emph{execution time}.

The number of iterations is a user-defined parameter for the algorithm.
A higher number of iterations indicates a more accurate score, but eventually the score converges to a stable value, with additional iterations contributing little.
We collect results up to a maximum of 50 iterations.
The PageRank~\cite{pagerank} paper reports that for a graph of 161M edges, 45 iterations are sufficient, while the graph used here only has 47M edges.

\subsubsection*{Results}

The results (\cref{fig:cross-opt-filter} and \cref{fig:cross-opt-aggregate}) validate two of our design principles: \textit{integration} (preprocessing expressed via query composition) and \textit{optimizability} (cross-optimization between queries and algorithms).
\graphalg{}/AvantGraph performs better than the other systems within a reasonable setting for the number of iterations (4-50).

For the `No self references' case, \graphalg{}/AvantGraph can push the query-side filter down to the edge table scan inside the algorithm thanks to the unified IR (\cref{sec:unified-ir}).
This makes it much faster than Neo4j, which needs to complete an expensive loading step before it can start executing the algorithm.
Even at 50 iterations, \graphalg{}/AvantGraph finishes execution before Neo4j has finished preprocessing the graph.
Query plan analysis indicates that DuckDB is also able to push the filter down to a table scan, but as shown in \cref{tab:proc-time}, it does not execute the PageRank algorithm as efficiently as \graphalg{}/AvantGraph.

For the `No duplicate edges' case, \graphalg{}/AvantGraph applies the loop-invariant code motion optimization (\cref{sec:licm}) and computes the aggregation only once.
This makes it scale better than DuckDB, which recomputes the aggregation every loop iteration.
Neo4j scales better than \graphalg{}/AvantGraph, but this only compensates for the high preprocessing overhead after 50 iterations, which is unrealistically high for this graph.

These preprocessing examples represent just the tip of the iceberg for cross-optimization opportunities.
The unified IR enables many other optimizations across the query-algorithm boundary: joining algorithm results with external tables, composing multiple algorithms in a pipeline, applying complex transformations to intermediate results, or dynamically selecting subgraphs based on algorithm outputs.
Each of these scenarios benefits from the same principle, treating queries and algorithms as first-class citizens in a single optimization framework, demonstrating the broad applicability of our approach beyond the specific cases studied here.

\section{Conclusions and Future Work}

We presented \graphalg{}, a domain-specific language for graph algorithms designed as a first-class citizen in graph database systems.
Grounded in linear algebra through MATLANG and for-MATLANG, \graphalg{} provides composable operations, clear semantics, and straightforward translation to an extension of relational algebra.

Integration of \graphalg{} with AvantGraph demonstrates that graph algorithms and graph queries can coexist in a unified high-performance framework.
By compiling \graphalg{} to the same internal representation as Cypher, we enable cross-optimization between queries and algorithms, eliminating data wrangling overhead.
Our experimental evaluation validates all four design principles: \textit{expressiveness} through diverse LDBC Graphalytics implementations; \textit{user-friendliness} via 2–10X code reduction; \textit{integration} through unified IPR representation; and \textit{optimizability} via sparsity analysis, loop-invariant code motion, and in-place aggregation.
Importantly, \graphalg{}/AvantGraph achieves \textit{superior performance} on critical algorithms like PageRank, SSSP, and WCC, matching or exceeding highly-optimized analytical databases while operating directly on persistent graph structures.

Our future work includes user studies futher to validate usability benefits, addressing AvantGraph's performance bottlenecks, and exploring additional compilation targets.

\begin{acks}
  This work has received funding from the European Union's Horizon Europe framework programme under grant agreement No. 101058573 as part of the SciLake project.
\end{acks}

\bibliographystyle{ACM-Reference-Format}

% Update style for Dutch naming
\DeclareRobustCommand{\VAN}[3]{#3}

\bibliography{misc,zotero}

\end{document}

%% file: loc-comparison.tex
\begin{tikzpicture}
\begin{axis}[
    ybar,
    bar width=8pt,
    height=5cm,
    enlargelimits=0.15,
    legend style={at={(0.5,-0.18)},
        anchor=north,legend columns=-1},
    ylabel={Lines of code},
    symbolic x coords={BFS,CDLP,PR,SSSP,WCC},
    xtick=data,
    nodes near coords,
    nodes near coords align={vertical},
    nodes near coords style={font=\tiny},
]
\addplot[fill=red] coordinates {
    (BFS,16)
    (CDLP,24)
    (PR,22)
    (SSSP,8)
    (WCC,13)
};
\addlegendentry{GraphAlg}
\addplot[fill=yellow] coordinates {
    (BFS,38)
    (CDLP,24)
    (PR,47)
    (SSSP,45)
    (WCC,30)
};
\addlegendentry{SQL/Python}
\addplot[fill=blue] coordinates {
    (BFS,72)
    (CDLP,90)
    (PR,103)
    (SSSP,76)
    (WCC,68)
};
\addlegendentry{Pregel/Java}
\end{axis}
\end{tikzpicture}

%% file: pr-body.tex
% AvantGraph
\noindent\begin{minipage}{.32\textwidth}
\lstset{language=GraphAlg}
\begin{lstlisting}[caption=GraphAlg (AvantGraph),frame=tlrb,showlines=true]{Name}
sink_pr = Vector<real>(n);
sink_pr<sinks> = pr;
redist = damping / cast<real>(n)
redist = redist * reduce(sink_pr);








w = pr (./) d;
pr[:] = teleport + redist;
pr += cast<real>(graph).T * w;




















\end{lstlisting}
\end{minipage}\hfill
% SQL
\begin{minipage}{.32\textwidth}
\lstset{language=SQLInPython, showstringspaces=false}
\begin{lstlisting}[caption={SQL/Python (DuckDB)},frame=tlrb]{Name}
(redist,) = conn.sql(f"""
  SELECT coalesce(
    ? * SUM(pr{i-1}.value), 0)
  FROM dangling
  JOIN PR{i-1} USING (id)
""", 
  [redistFactor]).fetchone()





conn.execute(f"""
  CREATE TABLE PR{i} AS
  SELECT
    g.v.id AS id,
    ($teleport + 
      ($damping * SUM(
        coalesce(pr{i-1}.value 
          / out_degree.degree, 0))) 
      + $redist) AS value
  FROM g.v
  LEFT JOIN g.e ON (
    g.e.target = g.v.id)
  LEFT JOIN pr{i-1} ON (
    pr{i-1}.id = g.e.source)
  LEFT JOIN out_degree ON (
    pr{i-1}.id = out_degree.id)
  GROUP BY g.v.id
""", {
  "teleport": teleport,
  "damping": damping,
  "redist": redist,
})
conn.execute(f"DROP TABLE pr{i-1}")
\end{lstlisting}
\end{minipage}\hfill
% Pregel
\begin{minipage}{.32\textwidth}
\lstset{language=Java, keywordstyle = \color{blue}, morekeywords={var}}
\begin{lstlisting}[caption={Pregel/Java (Neo4J)},frame=tlrb,showlines=true]{Name}
var sum = new MutableDouble();
sinks.forEach(sink -> {
  double pr = 
    context.doubleNodeValue(
      sink, PAGE_RANK);
  sum.add(pr);
});
redist = (
  context.config().damping() 
    / context.nodeCount()) 
      * sum.doubleValue();

double imp = 0;
for (var message : messages) {
  imp += message;
}
var damping = context.config().dampingFactor();
var teleport = (1 - damping) 
  / context.nodeCount();
newRank = teleport + (damping * imp) 
  + redist;
context.setNodeValue(PAGE_RANK, newRank);
if (!sink) {
  context.sendToNeighbors(newRank / context.degree());
}







\end{lstlisting}
\end{minipage}

%% file: table.tex
\begin{tabular}{l|ccc|ccc|ccc|ccc|ccc}
\toprule
Dataset & \multicolumn{3}{c}{BFS} & \multicolumn{3}{c}{CDLP} & \multicolumn{3}{c}{PR} & \multicolumn{3}{c}{SSSP} & \multicolumn{3}{c}{WCC} \\
 & GAA & DDB & N4J & GAA & DDB & N4J & GAA & DDB & N4J & GAA & DDB & N4J & GAA & DDB & N4J \\
\midrule
cit{-}Patents & \underline{0.69} & \textbf{0.43} & 4.67 & 62.18 & \underline{12.41} & \textbf{0.96} & \textbf{4.49} & \underline{5.73} & 7.59 &  \textit{\tiny N/A}  &  \textit{\tiny N/A}  &  \textit{\tiny N/A}  & \textbf{5.84} & \underline{14.31} & 17.96 \\
dota{-}league & \textbf{0.33} & \underline{0.43} & 3.03 & 35.11 & \textbf{9.26} & \underline{22.15} & \underline{8.66} & \textbf{7.61} & 15.37 & \underline{6.64} & \textbf{5.40} & 20.75 & \textbf{1.27} & \underline{2.34} & 7.56 \\
graph500{-}22 & \textbf{1.42} & \underline{1.44} & 11.78 & 97.24 & \textbf{18.66} & \underline{54.07} & \textbf{15.37} & 22.53 & \underline{18.71} &  \textit{\tiny N/A}  &  \textit{\tiny N/A}  &  \textit{\tiny N/A}  & \textbf{4.35} & \underline{11.15} & 23.22 \\
kgs & \underline{0.78} & \textbf{0.43} & 2.13 & 27.42 & \textbf{5.52} & \underline{17.55} & \textbf{4.09} & \underline{5.00} & 5.88 & \textbf{3.96} & \underline{4.62} & 11.04 & \textbf{1.22} & \underline{3.90} & 6.14 \\
wiki{-}Talk & \underline{0.63} & \textbf{0.33} & 1.97 & 17.41 & \underline{3.08} & \textbf{0.41} & \textbf{2.30} & \underline{2.41} & 5.73 &  \textit{\tiny N/A}  &  \textit{\tiny N/A}  &  \textit{\tiny N/A}  & \textbf{1.45} & \underline{1.90} & 5.51 \\
datagen{-}7\_5{-}fb & \textbf{0.52} & \underline{0.52} & 3.56 & 85.86 & \textbf{17.88} & \underline{25.61} & \textbf{7.00} & \underline{9.58} & 11.49 & \textbf{1.57} & \underline{1.78} & 9.05 & \textbf{1.41} & \underline{4.24} & 12.04 \\
datagen{-}7\_6{-}fb & \textbf{0.60} & \underline{0.72} & 4.23 & 106.69 & \textbf{22.76} & \underline{34.21} & \textbf{8.78} & 12.34 & \underline{9.95} & \textbf{1.86} & \underline{2.15} & 11.93 & \textbf{1.68} & \underline{5.33} & 14.40 \\
datagen{-}7\_7{-}zf & 14.84 & \textbf{3.35} & \underline{13.81} & \underline{130.33} & \textbf{28.40} & {\tiny OOM} & \textbf{16.59} & \underline{19.75} & 34.41 & \textbf{23.40} & \underline{32.25} & {\tiny OOM} & \textbf{16.41} & \underline{41.30} & {\tiny OOM} \\
datagen{-}7\_8{-}zf & \underline{22.64} & \textbf{4.75} & 23.23 & \underline{174.63} & \textbf{36.93} & {\tiny OOM} & \textbf{20.44} & \underline{25.67} & 35.79 & \textbf{34.97} & \underline{51.12} & {\tiny OOM} & \textbf{25.65} & \underline{62.07} & {\tiny OOM} \\
datagen{-}7\_9{-}fb & \textbf{1.20} & \underline{1.38} & 10.09 & 229.05 & \textbf{48.81} & \underline{80.59} & \textbf{19.30} & 26.26 & \underline{20.27} & \textbf{3.15} & \underline{3.90} & 25.80 & \textbf{3.96} & \underline{11.61} & 33.85 \\
\bottomrule
\end{tabular}

%% file: execution_times_filter.tex
\begin{tikzpicture}
\begin{axis}[
    height=6cm,
    width=8.5cm,
    xlabel={Number of Iterations},
    ylabel={Execution Time (s)},
    legend style={at={(0.02,0.98)},
        anchor=north west},
    grid=major,
    mark size=1pt,
]
\addplot coordinates {
    (1,3.112)
    (2,3.470)
    (3,3.828)
    (4,4.187)
    (5,5.119)
    (6,5.432)
    (7,5.795)
    (8,5.960)
    (9,5.973)
    (10,6.438)
    (11,7.072)
    (12,7.405)
    (13,7.422)
    (14,7.759)
    (15,9.036)
    (16,9.803)
    (17,9.523)
    (18,9.196)
    (19,10.877)
    (20,10.487)
    (21,10.685)
    (22,11.206)
    (23,12.921)
    (24,11.655)
    (25,12.255)
    (26,13.133)
    (27,13.882)
    (28,12.755)
    (29,13.128)
    (30,15.053)
    (31,15.746)
    (32,16.167)
    (33,15.567)
    (34,16.888)
    (35,17.597)
    (36,17.928)
    (37,16.780)
    (38,17.925)
    (39,18.971)
    (40,18.800)
    (41,18.743)
    (42,20.346)
    (43,21.472)
    (44,21.806)
    (45,22.255)
    (46,22.552)
    (47,22.281)
    (48,22.118)
    (49,22.593)
    (50,24.608)
};
\addlegendentry{AvantGraph}
\addplot coordinates {
    (1,2.036)
    (2,3.700)
    (3,5.644)
    (4,7.481)
    (5,8.617)
    (6,11.173)
    (7,12.912)
    (8,14.701)
    (9,15.701)
    (10,17.929)
    (11,19.305)
    (12,21.490)
    (13,22.877)
    (14,24.857)
    (15,26.620)
    (16,28.270)
    (17,29.827)
    (18,31.897)
    (19,33.876)
    (20,35.562)
    (21,37.433)
    (22,37.990)
    (23,40.771)
    (24,42.286)
    (25,44.187)
    (26,45.191)
    (27,47.089)
    (28,49.005)
    (29,50.565)
    (30,52.159)
    (31,53.968)
    (32,55.459)
    (33,57.562)
    (34,59.773)
    (35,60.976)
    (36,62.370)
    (37,64.202)
    (38,66.118)
    (39,68.046)
    (40,69.636)
    (41,71.533)
    (42,73.116)
    (43,75.311)
    (44,77.129)
    (45,78.670)
    (46,80.031)
    (47,81.837)
    (48,83.702)
    (49,85.294)
    (50,86.635)
};
\addlegendentry{DuckDB}
\addplot coordinates {
    (1,43.282)
    (2,43.084)
    (3,45.896)
    (4,46.148)
    (5,52.247)
    (6,46.433)
    (7,47.353)
    (8,47.292)
    (9,47.934)
    (10,48.776)
    (11,48.807)
    (12,49.548)
    (13,50.541)
    (14,49.980)
    (15,52.065)
    (16,56.717)
    (17,52.410)
    (18,52.508)
    (19,53.158)
    (20,52.468)
    (21,53.015)
    (22,53.717)
    (23,55.005)
    (24,55.599)
    (25,61.372)
    (26,60.864)
    (27,55.793)
    (28,55.920)
    (29,57.844)
    (30,56.498)
    (31,62.828)
    (32,57.516)
    (33,64.684)
    (34,58.408)
    (35,59.708)
    (36,61.641)
    (37,67.871)
    (38,58.910)
    (39,64.838)
    (40,60.656)
    (41,61.367)
    (42,63.668)
    (43,62.981)
    (44,61.014)
    (45,67.236)
    (46,62.519)
    (47,65.877)
    (48,63.135)
    (49,72.837)
    (50,67.178)
};
\addlegendentry{Neo4j}
\end{axis}
\end{tikzpicture}

%% file: execution_times_aggregate.tex
\begin{tikzpicture}
\begin{axis}[
    height=6cm,
    width=8.5cm,
    xlabel={Number of Iterations},
    ylabel={Execution Time (s)},
    legend style={at={(0.02,0.98)},
        anchor=north west},
    grid=major,
    mark size=1pt,
]
\addplot coordinates {
    (1,11.568)
    (2,12.257)
    (3,13.307)
    (4,14.525)
    (5,15.746)
    (6,16.836)
    (7,18.074)
    (8,19.333)
    (9,20.561)
    (10,21.595)
    (11,22.919)
    (12,24.061)
    (13,25.174)
    (14,26.601)
    (15,27.347)
    (16,28.620)
    (17,29.801)
    (18,30.774)
    (19,31.996)
    (20,33.407)
    (21,34.162)
    (22,35.501)
    (23,36.560)
    (24,37.769)
    (25,39.005)
    (26,40.082)
    (27,41.206)
    (28,42.295)
    (29,43.589)
    (30,44.792)
    (31,45.839)
    (32,47.056)
    (33,48.053)
    (34,49.120)
    (35,50.302)
    (36,51.508)
    (37,52.581)
    (38,53.730)
    (39,54.854)
    (40,56.285)
    (41,57.225)
    (42,58.427)
    (43,59.503)
    (44,60.651)
    (45,61.822)
    (46,62.838)
    (47,64.091)
    (48,65.281)
    (49,66.808)
    (50,67.678)
};
\addlegendentry{AvantGraph}
\addplot coordinates {
    (1,5.296)
    (2,8.413)
    (3,11.883)
    (4,14.907)
    (5,18.547)
    (6,21.326)
    (7,24.974)
    (8,27.984)
    (9,31.126)
    (10,34.106)
    (11,37.816)
    (12,41.135)
    (13,43.913)
    (14,47.139)
    (15,50.437)
    (16,53.718)
    (17,56.773)
    (18,60.188)
    (19,63.150)
    (20,66.330)
    (21,70.257)
    (22,73.250)
    (23,75.670)
    (24,79.415)
    (25,82.420)
    (26,85.775)
    (27,89.324)
    (28,92.287)
    (29,94.948)
    (30,98.664)
    (31,101.798)
    (32,105.216)
    (33,108.104)
    (34,110.931)
    (35,114.306)
    (36,117.789)
    (37,121.632)
    (38,123.702)
    (39,126.761)
    (40,130.994)
    (41,133.882)
    (42,137.046)
    (43,140.071)
    (44,143.289)
    (45,146.489)
    (46,149.194)
    (47,152.630)
    (48,156.012)
    (49,159.516)
    (50,162.538)
};
\addlegendentry{DuckDB}
\addplot coordinates {
    (1,41.650)
    (2,41.210)
    (3,44.373)
    (4,44.350)
    (5,44.688)
    (6,44.828)
    (7,47.007)
    (8,47.427)
    (9,46.320)
    (10,51.344)
    (11,48.680)
    (12,47.233)
    (13,48.540)
    (14,58.056)
    (15,52.236)
    (16,48.738)
    (17,51.359)
    (18,49.799)
    (19,56.560)
    (20,51.866)
    (21,52.296)
    (22,54.052)
    (23,56.191)
    (24,55.422)
    (25,56.734)
    (26,56.169)
    (27,55.393)
    (28,57.156)
    (29,62.125)
    (30,57.493)
    (31,62.124)
    (32,59.346)
    (33,60.143)
    (34,60.023)
    (35,61.108)
    (36,59.047)
    (37,58.517)
    (38,59.032)
    (39,62.153)
    (40,58.388)
    (41,61.832)
    (42,60.445)
    (43,62.178)
    (44,61.614)
    (45,63.275)
    (46,67.110)
    (47,67.775)
    (48,68.599)
    (49,67.981)
    (50,66.063)
};
\addlegendentry{Neo4j}
\end{axis}
\end{tikzpicture}